\documentclass{aastex61}

\usepackage{longtable}

\received{\today}
\submitjournal{ApJ}

\shorttitle{Warm CO gas in a low-mass dense core}
\shortauthors{Tokuda et al.}

\begin{document}

\title{Warm CO gas generated by possible turbulent shocks in a low-mass star-forming dense core in Taurus}

\correspondingauthor{Kazuki Tokuda}
\email{tokuda@p.s.osakafu-u.ac.jp}

\author[0000-0002-2062-1600]{Kazuki Tokuda}
\affiliation{Department of Physical Science, Graduate School of Science, Osaka Prefecture University, 1-1 Gakuen-cho, Naka-ku, Sakai, Osaka 599-8531, Japan}
\affiliation{Chile Observatory, National Astronomical Observatory of Japan, National Institutes of Natural Science, 2-21-1 Osawa, Mitaka, Tokyo 181-8588, Japan}

\author{Toshikazu Onishi}
\affiliation{Department of Physical Science, Graduate School of Science, Osaka Prefecture University, 1-1 Gakuen-cho, Naka-ku, Sakai, Osaka 599-8531, Japan}

\author{Kazuya Saigo}
\affiliation{Chile Observatory, National Astronomical Observatory of Japan, National Institutes of Natural Science, 2-21-1 Osawa, Mitaka, Tokyo 181-8588, Japan}

\author{Tomoaki Matsumoto}
\affiliation{Faculty of Sustainability Studies, Hosei University, Fujimi, Chiyoda-ku, Tokyo 102-8160, Japan}

\author{Tsuyoshi Inoue}
\affiliation{Department of Physics, Nagoya University, Chikusa-ku, Nagoya 464-8602, Japan}

\author{Shu-ichiro Inutsuka}
\affiliation{Department of Physics, Nagoya University, Chikusa-ku, Nagoya 464-8602, Japan}

\author{Yasuo Fukui}
\affiliation{Department of Physics, Nagoya University, Chikusa-ku, Nagoya 464-8602, Japan}

\author{Masahiro N. Machida}
\affiliation{Department of Earth and Planetary Sciences, Faculty of Sciences, Kyushu University, Nishi-ku, Fukuoka 819-0395, Japan}

\author{Kengo Tomida}
\affiliation{Department of Earth and Space Science, Osaka University, Toyonaka, Osaka 560-0043, Japan}

\author{Takashi Hosokawa}
\affiliation{Department of Physics, Kyoto University, Sakyo-ku, Kyoto 606-8502, Japan}


\author{Akiko Kawamura}
\affiliation{Chile Observatory, National Astronomical Observatory of Japan, National Institutes of Natural Science, 2-21-1 Osawa, Mitaka, Tokyo 181-8588, Japan}

\author{Kengo Tachihara}
\affiliation{Department of Physics, Nagoya University, Chikusa-ku, Nagoya 464-8602, Japan}



\begin{abstract}
We report ALMA Cycle 3 observations in CO isotopes toward a dense core, MC27/L1521F in Taurus, which is considered to be at an early stage of multiple star formation in a turbulent environment. Although most of the high-density parts of this core are considered to be as cold as $\sim$10 K, high-angular resolution ($\sim$20 au) observations in $^{12}$CO ($J$ = 3--2) revealed complex warm ($>$15--60 K) filamentary/clumpy structures with the sizes from a few tens of au to $\sim$1,000 au.
The interferometric observations of $^{13}$CO and C$^{18}$O show that the densest part with arc-like morphologies associated with the previously identified protostar and condensations are slightly redshifted from the systemic velocity of the core.  
We suggest that the warm CO clouds may be consequences of shock heating induced by interactions among the different density/velocity components that originated from the turbulent motions in the core.
However, such a small-scale and fast turbulent motion does not correspond to a simple extension of the line-width-size relation (i.e., Larson'{}s law), and thus the actual origin remains to be studied.
The high-angular resolution CO observations are expected to be essential in detecting small-scale turbulent motions in dense cores and to investigate protostar formation therein.
\end{abstract}


\section{Introduction} \label{sec:intro}
Interstellar turbulence is supposed to be one of the most important factors to regulate star formation activities. Shocks induced by supersonic turbulence dramatically increase the density and temperature in the post-shock layer and promote the structure formation, such as dense cores and protostars \citep[e.g.,][]{Padoan95,Padoan02,Klessen05,Matsumoto15b,Inoue18}.
\ Observational studies have been attempting to search for dense cores that originated from turbulent phenomena. Recent systematic surveys with ALMA found that there are no or a few starless cores with their internal substructures originated from turbulent fragmentation \citep[c.f.,][]{Padoan02,Offner10} in each star-forming region \citep{Dunham16,Kirk17}.
On the contrary, complex spatial/velocity gas structures in the protostellar envelopes with the spatial scale from 0.1 pc down to a few tens of au are also revealed both by the single-dish observations and the interferometric observations \citep[e.g.,][]{Tobin11,Tokuda14,Tokuda16,Maureira17}. 
Some multiscale polarization observations and numerical simulations suggest that turbulent motions, rather than magnetic field, are dynamically important  to form complex gas morphologies of protostellar envelopes \citep{Hull17a,Hull17b}. To understand such diversities/complexities at early stages of star formation, it is important to investigate physical properties of structures originating from the turbulent gas kinematics. In early phases of star formation, the shock waves can be formed by interactions among different density gas with the different velocities because the gas motions, such as infalling/outflowing gas, are supersonic. Detections of high-$J$ transitions of molecular lines (e.g., CO) excited by magnetohydrodynamic (MHD) shocks have been expected by theoretical modelings \citep[e.g.,][]{Pon12,Lehmann16}. Recent submillimeter observations \citep[e.g.,][]{van Kempen09a,van Kempen09b,Shinnaga09,Kristensen13} have detected warm envelopes seen in high-$J$ CO lines from protostellar sources. However, the origin and heating mechanisms of warm gas in low-mass star-forming dense cores are still under debate. For example, \cite{van Kempen09b} suggested several origins to produce the high-$J$ CO lines; (1) inner envelope heated by protostellar luminosity, (2) shocked gas in the outflow, and (3) quiescent gas heated by UV photons. \cite{Shinnaga09} also detected the extended CO ($J$ = 6--5, 7--6) emission with the size scale of a few thousand au in our current target, MC27/L1521F (see also later descriptions in this section). They argued that the warm gas may be coming from shock regions created by interactions between the collapsing envelopes and the internal disk-like materials around the protostar. Although their observations were not able to resolve the internal substructures of the dense core due to the lack of the angular resolution, they pointed out the importance for investigating such warm gas formed in an early phase of star formation to understand the evolution from dense cores to protostars.\\
\ We introduce our target object, MC27/L1521F \citep[e.g.,][]{Mizuno94,Onishi96,Onishi98,Onishi99,Onishi02,Codella97}, which is one of the densest cores among nearby low-mass star-forming regions, and it contains a very low-luminosity ($L_{\rm int}$$<$0.07 $L_{\odot}$) protostar \citep{Bourke06,Terebey09}. Earlier studies based on single-dish observations showed both the gas temperature derived from molecular line observations \citep[e.g.,][]{Codella97,Tatematsu04} and the dust temperature obtained from multiband dust continuum observations \citep[e.g.,][]{Kirk07,Sadavoy18} are as cold as $\sim$10 K, indicating that most of the gas and dust contents at the line of site are in the cold environment. The high-deuterium fractionation estimated by the N$_2$D$^+$/N$_2$H$^+$ ratio also supported the cold and evolved nature of the core \citep{Crapsi04,Crapsi05}. On the other hand, submillimeter single-dish observations detected CO ($J$ = 6--5, 7--6) lines, which trace the warm ($\sim$30--70 K) and dense ($\sim$10$^5$ cm$^{-3}$) component at the center of the core \citep{Shinnaga09}. Although the filling factor of this warm gas relative to the entire core is quite small, these facts demonstrated that multiple temperature mediums are coexisting within the core. \cite{Tokuda14} (hereafter, Paper I), and \cite{Tokuda16} (hereafter, Paper II) observed this core with ALMA and demonstrated the dynamical nature in this system. We found a few starless high-density cores, one of which has a very high density of $\sim$10$^{7}$ cm$^{-3}$ (MMS-2), within a region of several hundred au around a very low-luminosity protostar (MMS-1) detected by {\it Spitzer} \citep{Bourke06,Terebey09}. The molecular line observation showed several cores with arc-like structures, possibly due to the dynamical gas interaction. Similar arc-like structures have also been reproduced by hydrodynamical simulations in both with and without a magnetic field \citep{Matsumoto15a,Matsumoto17}, including the different mechanisms. 
These complex velocity/spatial structures indicate that in MC27/L1521F the turbulence may play an essential role in undergoing fragmentation in the central part of the cloud core, which is different from the classic scenarios of fragmentation in massive disks \citep{Larson87,Boss02,Machida08}. 
More recently, \cite{Tokuda17} (hereafter, Paper III) found that the central very low-luminosity protostar has the dynamical mass of $\sim$0.2 $M_{\odot}$ and the associated disk is extremely compact with the disk radius of $\sim$10 au with an indication of detachment nature from the surrounding dense environment. Paper III mainly focused on gas/dust distributions of the central protostar (MMS-1) seen in the 0.87 mm continuum and $^{12}$CO ($J$ = 3--2) observations to determine its evolutionary stage. In this paper, we investigate the properties of the surrounding gas around the central protostar to better understand the dynamical nature of this system by using the high-resolution ($\sim$0.2\arcsec) $^{12}$CO ($J$= 3--2) observations as well as $^{13}$CO ($J$ = 2--1) and C$^{18}$O ($J$ = 2--1) observations with a moderate angular resolution, $\sim$1\arcsec. 

\section{Observations} \label{sec:Obs}
We carried out ALMA Cycle 3 Band 7 (0.87 mm) and Band 6 (1.3 mm) observations with both continuum and molecular lines toward MC27/L1521F with a center position of ($\alpha_{J2000.0}$, $\delta_{J2000.0}$) = (4$^{\rm h}$28$^{\rm m}$38\fs996, +26\arcdeg51\arcsec35\farcs0). The observation settings of the Band 7 with the ALMA main array (the 12 m array) were shown in Paper III.  We also used the Atacama Compact Array (also know as the Morita array, the 7m array + the total power (TP) array) data, which were obtained in our previous program (Paper II), to recover the extended emission. The observation parameters of the Band 7 are summarized in Table \ref{tableObs} (the detailed descriptions were written in Papers II, and III).
 The Band 6 observations were made by the ALMA 12 m array, 7 m array, and TP array (single-dish) observations. The observations were carried out during 2016 March and September. There were three spectral windows targeting $^{13}$CO ($J$ = 2--1), C$^{18}$O ($J$ = 2--1), and N$_2$D$^{+}$ ($J$ = 3--2) and the correlator for each spectral window was set to have a bandwidth of 59 MHz with 1920 channels. Note that we used a spectral window for the observations of the continuum emission with a bandwidth of 2.0 GHz (15.6 MHz $\times$ 128 channels). The $uv$ ranges of the 7 m array and the 12 m array are 5.70--30.7 k$\lambda$ and 10.0--810 k$\lambda$, respectively. The data were reduced using the Common Astronomy Software Application (CASA) package \citep{McMullin07}. The original synthesized beam of the 12 m array data alone is 0\farcs57 $\times$ 0\farcs37 by using the briggs weighting with the robust parameter of 0.5. We applied additional $uv$ tapering of the visibilities to enhance the sensitivities. The resultant synthesized beams and the sensitivities are summarized in Table \ref{tableObs}. The missing fluxes against the TP array observations are also shown in Table \ref{tableObs}. 
The 12 m array observations have the large amounts of missing flux of the extended emission from the source (see also Sec. \ref{subsec:COfilament} and \ref{subsec:1318}). We combined the datasets (12 m + 7 m + TP) by using the feathering technique in the CASA instructions obtained by ALMA (see also Paper II). We note that the 1.3 mm continuum observations reproduce the previously obtained continuum images well at the wavelength of 1.2 mm and 0.87 mm (Papers I, and II).

\begin{deluxetable}
{lccccccccc}  
\tabletypesize{\scriptsize}
\tablecaption{Beam sizes, sensitivities, and missing flux \label{tableObs}}
\tablewidth{0pt}
\tablehead{
  & \multicolumn{3}{c}{Band 7} & \multicolumn{6}{|c}{Band 6}\\
  \hline
  & \multicolumn{3}{c}{$^{12}$CO ($J$ = 3--2)} & \multicolumn{3}{|c}{$^{13}$CO ($J$ = 2--1)} & \multicolumn{3}{|c}{C$^{18}$O ($J$ = 2--1)}\\
  \hline
Array & 12 m  & 7 m $^{\rm a}$ & TP $^{\rm a}$ & 12 m & 7 m & TP  & 12 m  & 7 m  & TP 
}
\startdata
Beam size & 0\farcs18 $\times$ 0\farcs11 & 5\farcs0 $\times$ 4\farcs1 & 19\arcsec & 1\farcs2 $\times$ 1\farcs1 & 9\farcs3 $\times$ 5\farcs5 & 30\arcsec & 1\farcs2 $\times$ 1\farcs1 & 8\farcs1 $\times$ 6\farcs0 & 30\arcsec \\
RMS noise level $^{\rm b}$ (mJy\,beam$^{-1}$)  & 1.7  &  43  & 560  & 16  & 78  & 290  & 15  & 55  & 330  \\
RMS noise level $^{\rm b}$ (K)  & 0.91  &  0.021  & 0.016  & 0.29  & 0.038  & 0.0084  & 0.30  & 0.029  & 0.0096  \\
Missing flux $^{\rm c}$  & $\sim$49 \%  & $\sim$42 \% &$\cdots$ & $\sim$92 \%  &  $\sim$86 \% & $\cdots$ & $\sim$92 \%  & $\sim$82 \% & $\cdots$ \\
\enddata
\tablenotetext{\rm a}{Previously obtained data from Paper II.}
\tablenotetext{\rm b}{The velocity resolutions of the Band 7 and the Band 6 are $\sim$0.85 km\,s$^{-1}$ and $\sim$0.1 km\,s$^{-1}$, respectively.}
\tablenotetext{\rm c}{Missing flux of each array relative to the total flux measured by the TP array observations in the area of the TP beam size at the field center.}
\end{deluxetable}

\section{Results \label{sec:results}} 
\subsection{CO filaments at the center of the dense core, MC27/L1521F, revealed at $\sim$20 au resolution\label{subsec:COfilament}}
\begin{figure}[htbp]
\includegraphics[width=180mm]{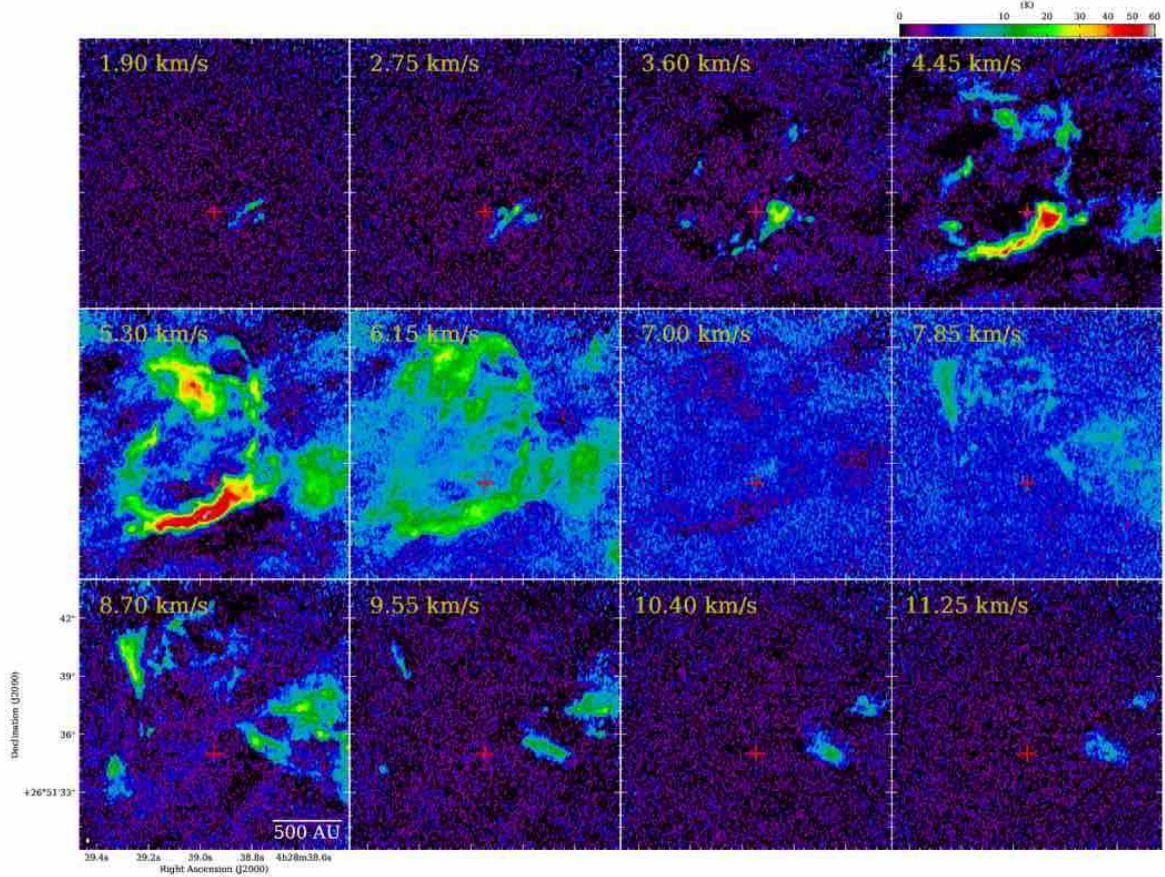}
\caption{Velocity-channel maps of the $^{12}$CO ($J$= 3--2) toward MC27/L1521F. Color-scale images show the brightness temperature maps by combining the 12 m, 7m, and TP array data. The central velocities are given in the upper left corner of each panel. The angular resolution, 0\farcs18 $\times$ 0\farcs11, is given by the white ellipse in the lower left corner of the bottom left panel. The red crosses in each panel denotes the position of the 0.87 mm continuum peak, ($\alpha_{J2000.0}$, $\delta_{J2000.0}$) = (4$^{\rm h}$28$^{\rm m}$38\fs9462, +26\arcdeg51\arcsec35\farcs004), which is the location of the protostar, MMS-1 (Paper III). \label{fig:chmapCO}}
\end{figure}

Figure \ref{fig:chmapCO} shows velocity-channel maps of $^{12}$CO ($J$ = 3--2) with a velocity resolution of 0.85 km\,s$^{-1}$. The angular resolution is 0\farcs18 $\times$ 0\farcs11, corresponding to $\sim$ 25 au $\times$ 15 au at a distance of the Taurus molecular cloud ($\sim$140 pc, \citealt{Elias78}).
The images are made by combining with the 12 m, 7 m, and TP array data (see Table \ref{tableObs}). 
The extended emissions around the systemic velocity of the core, 5.30--7.85 km\,s$^{-1}$ are mainly recovered by the TP array observations and the brightness temperature is $\lesssim$10 K. The missing flux around the systemic velocity and the high-velocity ($\le$4.45 km\,s$^{-1}$, and $\ge$8.70 km\,s$^{-1}$) components with the 12 m array are estimated to be $\gtrsim$60 \% and $\lesssim$10 \%, respectively. These facts demonstrate that the high-velocity blueshifted/redshifted components are mainly composed of the narrow filamentary or compact structures instead of the diffuse emission (see also a case of the $^{12}$CO ($J$= 2--1) in \cite{Takahashi13}).
We note that the total missing flux of the $^{12}$CO ($J$= 3--2) emission with the 12 m and the 7 m array is smaller than that of the $^{13}$CO and the C$^{18}$O ($J$= 2--1) (Table \ref{tableObs}). The $^{12}$CO observations, therefore, cannot trace the total gas mass at the line of sight due to the self-absorption around the systemic velocity of the core. 

\begin{figure}[htbp]
\plotone{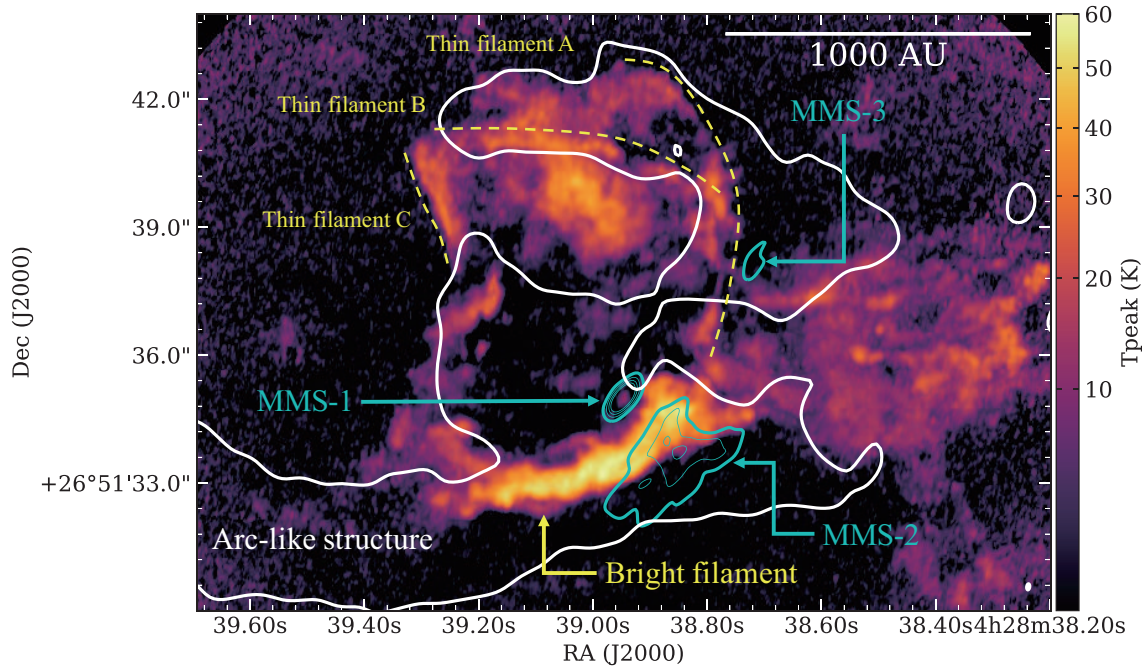}
\caption{$^{12}$CO ($J$ = 3--2) distributions and remarkable features at the center of the dense core, MC27/L1521F. Color-scale image shows the peak temperature maps of $^{12}$CO ($J$= 3--2) with the 12 m array data alone. The angular resolution, 0\farcs18 $\times$ 0\farcs11, is given by the ellipse in the lower right corner. White contour shows velocity-integrated intensity of HCO$^+$ ($J$ = 3--2) with a velocity range of 6.8--7.0 km\,s$^{-1}$ (Paper I). Cyan contours represent the image of 0.87 mm dust continuum (Paper II). \label{fig:bright}}
\end{figure}

We can see many filamentary clouds and compact structures in Figure \ref{fig:chmapCO}. We note that many of these prominent structures could not clearly be seen in our previous studies with a resolution of $\sim$0\farcs3 $\times$ 0\farcs7 in the same line (see also Figure 6 in Paper II).  The highest temperature part is located at southwest of the central protostar at the velocity ranges of 4.45--5.30 km\,s$^{-1}$. We call this high-temperature part as a $``$Bright filament''. Figure \ref{fig:bright} shows the $^{12}$CO ($J$= 3--2) peak temperature map within the velocity range as shown in Figure \ref{fig:chmapCO}. We also overlay some remarkable structures, high-density dust cores and complex arc-like structures (Paper I, and II), on the $^{12}$CO map. The physical properties are summarized in Table \ref{tableParams}. 
The maximum peak temperature in the CO ($J$= 3--2) is around 60 K, indicating that the kinematic temperature should also be high. The brightness temperature of $^{13}$CO ($J$ = 2--1) toward the Bright filament is $\sim$5 K at a resolution of $\sim$0\farcs5. We estimate the optical depth, the column density, and the kinematic temperature of Bright filament as follows. 
If we spatially smooth the angular resolution of the $^{12}$CO ($J$= 3--2) to that of the $^{13}$CO, $\sim$0\farcs5, the peak brightness temperature of the Bright filament is measured as $\sim$50 K. 
We assume that $^{12}$CO ($J$= 3--2) is optically thick and that the peak temperature is the same as that of $^{12}$CO ($J$= 2--1). We thus estimate the optical depth of $^{12}$CO, $\tau_{12}$, from the intensity ratio of $^{12}$CO ($J$= 3--2) and $^{13}$CO ($J$= 2--1) by using the following equation:
\begin{equation}
I_{\rm ^{12}CO}/I_{\rm ^{13}CO} = [1-{\rm exp}(-\tau_{12})]/[1-{\rm exp}(-\tau_{12}/R_{12/13})]\\
\end{equation}
where $I_{\rm ^{12}CO}$ and $I_{\rm ^{13}CO}$ are the peak brightness temperatures of $^{12}$CO($J$= 3--2) and $^{13}$CO($J$= 2--1), respectively, and $R_{12/13}$ is the intrinsic abundance ratio between $^{12}$CO and $^{13}$CO ($\sim$70, e.g., \citealt{Frerking82}). The $\tau_{12}$ is calculated as $\sim$8, indicating that the $^{12}$CO emissions at the Bright filament are optically thick. This value is also consistent with that derived from the earlier study \citep{Takahashi13}. Therefore, the peak brightness temperature of the Bright filament fairly reflects the kinematic temperature. The H$_2$ column density of the Bright filament is calculated to be $\sim$3 $\times$ 10$^{21}$ cm$^{-2}$ by assuming the fractional abundance of $^{13}$CO relative to H$_2$ of 1.4 $\times$ 10$^{-6}$ (e.g., \citealt{Frerking82}). The volume density derived by dividing the column density by the width of the filament, $\sim$300 au, is calculated to be $\sim$8 $\times$ 10$^5$ cm$^{-3}$. We also detected high-density tracers, such as HCN ($J$= 3--2) and CS ($J$ = 5--4), at the position of the Bright filament (see also Figure 2 in Paper I). These results clearly demonstrate that both the density and temperature of the Bright filament are actually high.\\

\begin{deluxetable}
{lcccccccc}  
\tabletypesize{\scriptsize}
\tablecaption{Physical parameters of remarkable objects in MC27/L1521F  \label{tableParams} denoted in Figure \ref{fig:bright}}
\tablewidth{0pt}
\tablehead{
Name & Length  &  $N$ & $n$  & $T_{\rm kin}$ & $dv^{\rm c}$ & Total Mass & Tracers & References \\
{} & (au)  &  (cm$^{-2}$) & (cm$^{-3}$)  & (K) & (km\,s$^{-1}$) & ($M_{\odot}$) & {} & {} }
\startdata
MC27/L1521F$^{\rm a}$   & $\sim$20000 &  $\sim$3 $\times$ 10$^{22}$  & 10$^{5}$--10$^{6}$  & $\sim$10  & 0.3--0.5           & $\sim$3--4                   & e.g., dust, H$^{13}$CO$^+$, N$_2$H$^+$ & [1,2,3,4] $^{\rm e}$ \\
MMS-1                 & $\sim$20       &  3 $\times$ 10$^{23}$             & $\cdots$                  & $\sim$100 & $\cdots^{\rm d}$ & 8 $\times$ 10$^{-5}$            & dust, $^{12}$CO & Paper I,II,III \\
MMS-2                 & 380--380       &  0.37--1.3 $\times$ 10$^{23}$ & $\sim$10$^{7}$      & $\sim$10   & 0.40                    & 1.5--3.9 $\times$ 10$^{-3}$  & dust, H$^{13}$CO$^{+}$, C$^{17}$O, C$^{18}$O & Paper I,II \\
MMS-3                 & 35--160         &  1.5--5.1 $\times$ 10$^{22}$   & $\sim$10$^{6}$      & $\sim$10   & 0.36                    & 1.8--4.2 $\times$ 10$^{-4}$  & dust, H$^{13}$CO$^{+}$, C$^{18}$O & Paper I,II \\
Arc-like structure & $\sim$2000   &  3 $\times$ 10$^{21}$              & $\sim$10$^{5}$      & $\sim$10  & 0.3                      &  $\sim$10$^{-4}$                   & HCO$^{+}$, C$^{18}$O, $^{13}$CO &Paper I, This work \\
Bright filament     & $\sim$1000   & 3 $\times$ 10$^{21}$               & $\sim$10$^{6}$      & $\sim$60  & 1.6                      & $\sim$10$^{-4}$                    & $^{12}$CO, $^{13}$CO, HCO$^{+}$, HCN, CS & Paper I, This work \\
Thin filaments$^{\rm b}$      &  $\sim$1000  & $\sim$10$^{21}$               & $\cdots$                 & $>$10--30      & $\lesssim$2               & $\sim$10$^{-5}$                    & $^{12}$CO & This work \\
Tiny CO clumps$^{\rm b}$   &  $\sim$30      & $\sim$10$^{21}$                     & $\cdots$                 & $>$15      & $<$0.85              & $\sim$10$^{-6}$                    & $^{12}$CO & This work \\
\enddata
\tablenotetext{\rm a} {Physical parameters of the parental core derived by single-dish studies.}
\tablenotetext{\rm b} {Typical parameters of the identified structures.}
\tablenotetext{\rm c} {Observed line width (FWHM) of averaged spectrum in each identified structures by fitting the profile with a single Gaussian profile.}
\tablenotetext{\rm d} {The gas motion is considered to be a Keplerian rotation with the central protostellar mass of $\sim$0.2 $M_{\odot}$ (Paper III).}
\tablenotetext{\rm e} {[1] \cite{Onishi99}, [2] \cite{Onishi02}, [3] \cite{Crapsi04}, [4] \cite{Tatematsu04}}
\end{deluxetable}

\begin{figure}[htbp]
\includegraphics[width=180mm]{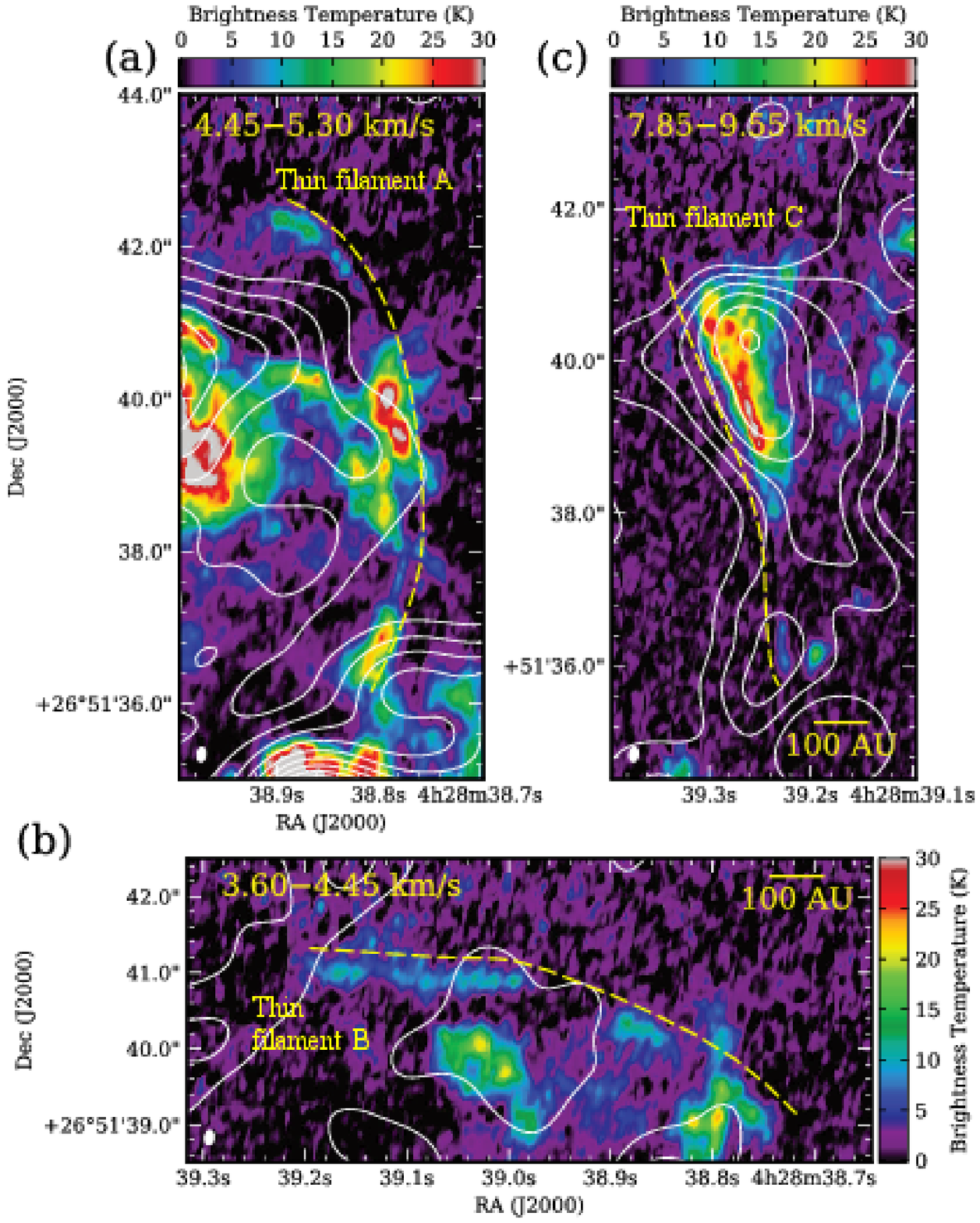}
\caption{
Magnified images of thin filaments. Thin filament A, B, and C in Figure \ref{fig:bright} are shown in panels (a), (b), and (c), respectively. The color-scale images show peak brightness temperature maps of the $^{12}$CO ($J$= 3--2) with the velocity ranges shown in the upper corners in each panel. The angular resolution, 0\farcs18 $\times$ 0\farcs11, is given by the white ellipse in the lower left corners. White contours show integrated intensity maps of the $^{13}$CO ($J$= 2--1) with the 12 m array alone. The lowest contour and the subsequent contour step are 0.06 and 0.15 K\,km\,s$^{-1}$, respectively.
\label{fig:filamentA}}
\end{figure}

Figures \ref{fig:chmapCO}, and \ref{fig:bright} demonstrate several very-narrow filamentary structures with length scale of $\sim$1,000 au. We call these structures as $``$thin filaments.'' We numbered three remarkable filaments identified by eye as shown in Figures \ref{fig:bright} and \ref{fig:filamentA}. The typical velocity width of the thin filaments is $\lesssim$2 km\,s$^{-1}$. 
Thin filament A is connecting to the Bright filament and Thin filament B both in the spatial and velocity space. The $^{13}$CO emission could not be detected toward Thin filament A with the present resolution and sensitivity. This fact indicates that the H$_2$ column densities are not as high as $\lesssim$10$^{21}$ cm$^{-2}$. 
Although we can identify the $^{13}$CO emission on Thin filament C, the peak position is shifted from the $^{12}$CO peak (Fig. \ref{fig:filamentA} (c)). 
The brightness temperatures of the filaments are $\sim$10--30 K in spite of the fact that the filament widths are similar to the beam size. It means that the actual intensities of the filamentary structures are considered to be higher than the observed values. The possible origins of the filamentary structures are discussed in Section \ref{subsec:originshock}. 

\subsection{Tiny CO clumps \label{subsec:tiny}}
\begin{figure}[htbp]
\includegraphics[width=180mm]{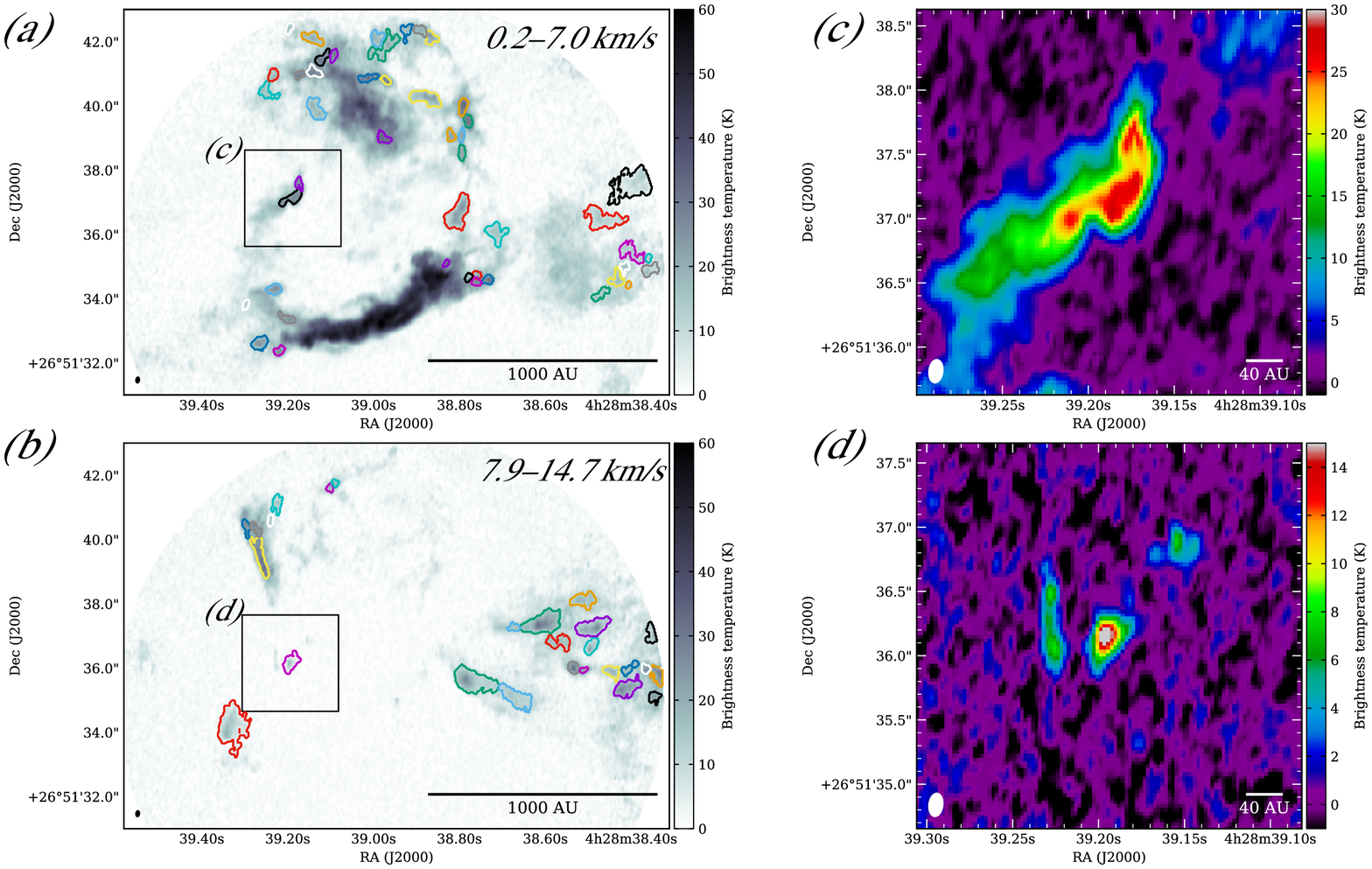}
\caption{Tiny CO clumps in MC27/L1521F. (a),(b) Grayscale images show the peak brightness temperature maps of $^{12}$CO ($J$ = 3--2) with the velocity ranges of 0.2--7.0 km\,s$^{-1}$ for (a), and 7.9--14.7 km\,s$^{-1}$ for (b), respectively. Color contours represent boundaries of the tiny CO clumps identified by the dendrogram analysis as $``$leaves'' (see the text). (c), (d) Magnified views of the tiny clumps as shown by rectangles in panels (a) and (b) are shown in color scale. The angular resolution, 0\farcs18 $\times$ 0\farcs11, is given by the white ellipse in the lower left corner of the bottom left panel.\label{fig:dendro}}
\end{figure}

Many tiny structures with size scales of a few tens of au are also observed in the $^{12}$CO ($J$= 3--2) channel maps of Figure \ref{fig:chmapCO}.
Some of them are isolated from surrounding gas, such as the filamentary structures as mentioned in Sec. \ref{subsec:COfilament} and the others have high-intensity contrast against the surrounding gas distributions. 
In order to characterize the properties of the tiny structures in the $^{12}$CO ($J$= 3--2) emission, we used a dendrogram algorithm (astrodendoro package, \citealt{Rosolowsky08}). The concept of the dendrograms is to decompose the structures of molecular gas as a $``$structure tree'' in a multidimensional data \citep{Houlahan92}. The structures were decomposed into three categories: the smallest structures (leaves), the largest continuous structures (trunks), and those intermediate in scale (branches). In this analysis, we extracted the leaves to investigate the nature of the tiny structures. We adopted the following criteria: the maximum and minimum peak brightness temperature at the leaves are 15 and 35 K, to exclude the weak components and Bright filament as shown in \ref{subsec:COfilament}. We refer to the identified leaves as $``$tiny CO clumps'' hereafter. The boundaries of the tiny CO clumps are shown on the $^{12}$CO distributions in Figure \ref{fig:dendro}. The total number of the clumps is 72. The typical radius is $\sim$25 au, showing that the tiny CO clumps are unresolved or marginally resolved even in the present high-angular resolution observations. 
The velocity dispersions cannot be derived in this analysis because most of the identified clumps have only a few pixels in the velocity space. This means that the velocity dispersions are smaller than the present velocity resolution, 0.85 km\,s$^{-1}$.
The physical properties in each clump are shown in Table \ref{table:COclumps} in the Appendix. Such extremely small structures are not considered to be self-gravitating objects because they are two orders of magnitude smaller than Jeans length with the gas density of the parental core ($\sim$10$^5$--10$^6$ cm$^{-3}$). We discuss possible origins of the tiny CO clumps in Sec. \ref{subsec:originshock}.

\subsection{Gas distributions in $^{13}$CO ($J$ = 2--1) and C$^{18}$O ($J$ = 2--1) \label{subsec:1318}}
\begin{figure}[htbp]
\includegraphics[width=180mm]{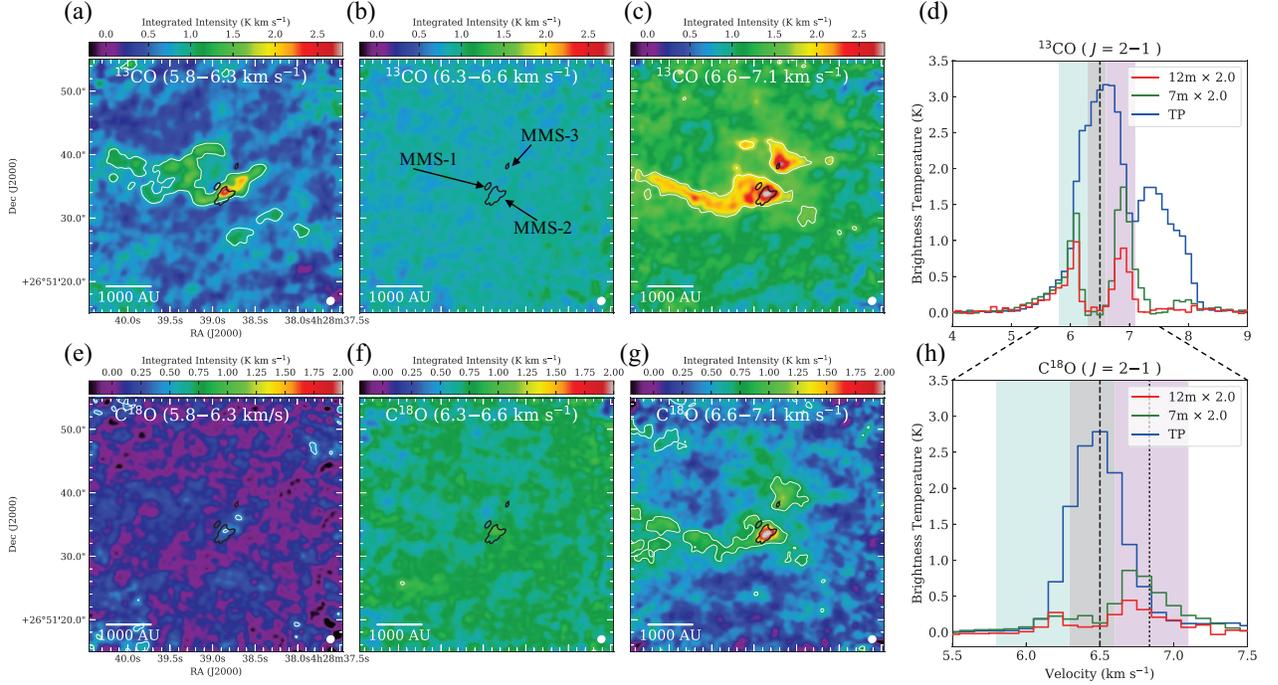}
\caption{Velocity-integrated images and averaged spectra of $^{13}$CO ($J$ = 2--1) and C$^{18}$O ($J$ = 2--1) in MC27/L1521F. (a)-(c), (e)-(g) Color-scale images show the $^{13}$CO and the C$^{18}$O maps combined with the 12 m, 7m, and TP array data. The integrated velocity ranges are shown in the upper side of each panel and panels (d) and (h). The angular resolutions of the combined data are given in the lower left corners. Boundaries of the 12 m array data alone are shown in white contours with the level of 5$\sigma$ in each molecule. The 1$\sigma$ levels of the $^{13}$CO and the C$^{18}$O are 7.5 $\times$ 10$^{-2}$ K km\,s$^{-1}$ and 6.6 $\times$ 10$^{-2}$ K km\,s$^{-1}$, respectively. Black contours represent 0.87 mm continuum emission, same as Figure \ref{fig:bright}. (d), (h) Red, green, and blue profiles show $^{13}$CO/C$^{18}$O averaged spectra obtained with each array over the regions inside circles with a radius of 30\arcsec (beam sizes of the TP array observations) at the position of MMS-1. Color hatches show the integrated velocity ranges in panel (a)-(c), and (e)-(g). Dashed lines denote the systemic velocity of the core, 6.5 km\,s$^{-1}$. The dotted line in panel (h) represents the centroid velocity of MMS-2 (see also Table \ref{table:velocity}). \label{fig:1318}}
\end{figure}

We present the envelope gas distributions seen in the $^{13}$CO and C$^{18}$O ($J$ = 2--1) observations. The combined ALMA images with the 12 m,  7 m, and TP arrays of the $^{13}$CO/C$^{18}$O, and the averaged spectra seen in each array are shown in Figure \ref{fig:1318}. We divide the intensity maps of the $^{13}$CO/C$^{18}$O into three velocity ranges: the blueshifted side (5.8--6.3 km\,s$^{-1}$), near the systemic velocity (6.3--6.6 km\,s$^{-1}$), and the redshifted side (6.6--7.1 km\,s$^{-1}$). The velocity ranges are also shown in color hatches of Figure \ref{fig:1318} (d), (h). The overall $^{13}$CO distributions traced by the 12 m array are quite similar to that of the HCO$^{+}$ ($J$= 3--2) (Papers I, and II). This means that both lines are roughly tracing the same gas expected for some local differences (see Sec. \ref{subsec:originshock}). 
One of the most striking features at the redshifted velocity ranges, arc-like structures, which are already shown in our previous observations (Figure \ref{fig:bright}, and see also Paper I) are also remarkable in the $^{13}$CO/C$^{18}$O observations. Under the assumption of the LTE condition and optically thin emission, we estimate the mass and column density from the $^{13}$CO observations with the 12 m array data alone, excluding the contributions from the extended emission. The physical parameters of the arc-like structure are shown in Table 2. There are significant missing fluxes around the systemic velocity of MC27/L1521F, 6.5 km\,s$^{-1}$ \citep[e.g.,][]{Onishi99} with the 12 m array and the 7m array (see also Table \ref{tableObs}). It means that the large amounts of extended gas traced by the $^{13}$CO/C$^{18}$O in this core are not seen in the interferometric observations. 
We investigate the positions of the high-density structures traced by the interferometer in the velocity axis. The central velocities of the blueshifted/redshifted gas of the C$^{18}$O, the previously identified dust condensations (MMS-2, 3), the protostellar source (MMS-1), and the parental core are listed in Table \ref{table:velocity}.  We find that the central velocities of MMS-1, -2 and -3 are slightly redshifted from that of the parental core, 6.5 km\,s$^{-1}$, and close to that of the redshifted component of the C$^{18}$O (Table \ref{table:velocity}).  The positions of MMS-2 and MMS-3, are consistent with the local peaks of the $^{13}$CO/C$^{18}$O at the redshifted side (Figs. \ref{fig:1318} (c), (g)). The intensities of the C$^{18}$O at the redshifted velocities are significantly stronger than that of the blueshifted side (Figs. \ref{fig:1318} (e) and (g)). In summary, these results represent that the high-density parts in MC27/L1521F seen in the interferometric observations are slightly redshifted from the systemic velocity of the parental core. 

\begin{deluxetable}
{lcccc}  
\tabletypesize{\scriptsize}
\tablecaption{Centroid velocities of structures traced by molecular lines \label{table:velocity}}
\tablewidth{0pt}
\tablehead{
Source & Centroid velocity (km\,s$^{-1}$) 
& Tracer & Instrument
}
\startdata
Parental core (MC27/L1521F) & 6.5            & H$^{13}$CO$^{+}$$^{(1)}$, N$_2$H$^{+}$$^{(2)}$, N$_2$D$^+$$^{(3)}$ & Single-dish\\ 
Redshifted component            & $\sim$6.7               & C$^{18}$O              & 12 m + 7 m array\\
Blueshifted component            & $\sim$5.0               & C$^{18}$O              & 12 m + 7 m array\\
MMS-1                                     & 6.75             & $^{12}$CO$^{(4)}$              & 12 m array\\
MMS-2                                     & 6.84               & H$^{13}$CO$^{+}$$^{(5)}$  & 12 m array\\
MMS-3                                     & 6.87               & H$^{13}$CO$^{+}$$^{(5)}$  & 12 m array\\
Diffuse gas                               & 7.4                 & $^{13}$CO                          & Single-dish\\
\enddata 
References \tablenotetext{}{$^{(1)}$\cite{Onishi99}, $^{(2)}$\cite{Tatematsu04}, $^{(3)}$\cite{Crapsi04}, $^{(4)}$Paper III, $^{(5)}$ Paper I}
\end{deluxetable}

The $^{13}$CO emission ($J$ = 2--1) of the TP array exhibits two velocity components at the systemic velocity of MC27/L1521F, 6.5 km\,s$^{-1}$, and 7.4 km\,s$^{-1}$. The 7.4 km\,s$^{-1}$ velocity component has already been detected in the previous $^{13}$CO ($J$ = 1--0) observations with single-dish telescopes \citep{Mizuno95,Goldsmith08,Narayanan08,Takakuwa11}. The absence of the strong emissions of the second velocity component with the interferometric observations shows that the emission is mainly originated from the extended gas. The systemic velocity component of MC27/L1521F corresponding to that of the L1521 molecular complex, while the 7.4 km\,s$^{-1}$ one is a part of the velocity component of the filamentary structure in B213 and L1495 \citep{Goldsmith08,Takakuwa11}. 

\section{Discussions \label{sec:dis}} 

\subsection{Possible origins of warm gas seen as Bright filament, thin filaments and tiny CO clumps\label{subsec:originshock}}
In this section, we discuss origins of the $^{12}$CO ($J$= 3--2) distributions seen as Bright filaments, the thin filaments and the tiny CO clumps. The brightness temperature of the structures mentioned above exceeds 10 K. We have to consider heating mechanisms to realize the warm CO gas in the cold dense core. 
There are several candidates to heat up the clouds; (1) protostellar heating from MMS-1 (L1521F-IRS), (2) shocks induced by the turbulent motions in the core, and (3) outflow activities in the past. The second one provides a convincing interpretation of the distributions of warm CO gas. There is no strong evidence to exclude the third possibility as well. We consider each of the mechanisms in the following paragraphs. 
\\
\ A possible mechanism to realize the warm cloud is radiation from the protostar. However, the observed low-luminosity cannot contribute such the heating to the size scale of 100$-$1,000 au. We assume that the temperature profile around the protostar follows the radiative equilibrium equation, $T \propto r^{-2/(4+\beta)}$, where $r$ is a distance from the protostar and $\beta$ is a dust opacity, $\sim$0.4 (Paper II). The maximum distances that can be heated up to $\sim$60 K due to the protostellar heating are limited to 10 au if we adopted a temperature at 1 au radius from the protostar, $T_{\rm 1au}$ = 180 K (Paper III). Moreover, the distributions of the Bright filament are localized to the southwest side of the protostar rather than isotropic. \cite{Shinnaga09} also suggested that the low-luminosity of the protostar is insufficient to pump CO molecules up to high-$J$ levels over a few thousand au. Thus, the protostellar heating to produce the warm gas is unlikely in this system. \\
\ We discuss whether the origins of the warm CO gas can be explained by the turbulent motions within the dense core. We calculated one-dimensional shock heating at the post-shock layer by using the Rankine-Hugoniot equations. 
Because the main coolants are dust particles, molecular line cooling is not considered in the simulation.
The energy exchange between the gas and the dust grain follows an Equation (57) in \cite{Leung75}. We set the gas density of 10$^5$ cm$^{-3}$ and the temperature of 10 K at the preshock layer, which are averaged properties at the central part of the core \citep[e.g.,][]{Onishi02,Crapsi04}. As a result, we found that the maximum temperature at the shock front becomes $\sim$200 K in the case of the downstream velocity of 2 km\,s$^{-1}$ with the e-holding time scale of $\sim$3,000 years.  
The temperature at a distance of $\sim$300 au from the shock front is $\sim$60 K. This is consistent with the observed properties of the Bright filament (Sec. \ref{subsec:COfilament}, see also calculations by K. Saigo et al. in prep.).
 \cite{Shinnaga09} also proposed that heating by shocks is a most likely mechanism to interpret the high-$J$ CO emission in this object. 
The observed line widths of HCO$^{+}$, CO, and $^{13}$CO with the single-dish beam size (\citealt{Onishi99,Takakuwa11,Takahashi13}, see also Figure \ref{fig:1318} in Sec \ref{subsec:1318}) are $>$1--2 km\,s$^{-1}$, which is much higher than the thermal line width, indicating that the gas motions with the density of 10$^3$--10$^5$ cm$^{-3}$ are considered to be supersonic. Numerical simulations by \cite{Matsumoto15a} also reproduced such a supersonic velocity field due to the gravitational interactions among the protostars (sink particles) and the envelope gas. We also illustrate the spatial relations among the brightness temperature distributions seen in $^{12}$CO ($J$= 3--2) and redshifted/blueshifted gas of HCO$^{+}$ ($J$= 3--2) as well as $^{13}$CO ($J$ = 2--1) in Figure \ref{fig:Comp}. The blueshifted and redshifted clouds have distributions that are complementary to each other (see also \citealt{Onishi15}), and the Bright filament and the thin filaments seem to be located at the interface layer between two velocity clouds. We note that similar complementary gas distributions are often seen on much larger scales at high-mass star-forming regions induced by colliding clouds, and these distributions are reproduced by many numerical simulations  \citep[e.g.,][]{Matsumoto15b,Torii17}.s
These distributions indicate that the different velocity components with a relative velocity of $\sim$1 km\,s$^{-1}$ at the line of sight are interacting with each other to create the shock-heated layer.  Although the overall gas distributions traced by the HCO$^{+}$ and $^{13}$CO observations are similar, there are small discrepancies locally possibly due to the shock-induced enhancement of the HCO$^+$ abundance \citep{Bachiller97,Jorgensen04}. Along with Thin filament A, there are no $^{13}$CO emissions except HCO$^+$ (Sec. \ref{subsec:COfilament}). The peak position of the $^{12}$CO on Thin filament C coincides with that of the HCO$^+$ rather than the $^{13}$CO in the same velocity ranges. In summary, we propose that the current $^{12}$CO ($J$= 3--2) distributions including the high-temperature gases can be explained by the shock phenomena induced by the present turbulent velocity field in this system.\\
\begin{figure}[htbp]
\label{fig:Comp}
\includegraphics[width=180mm]{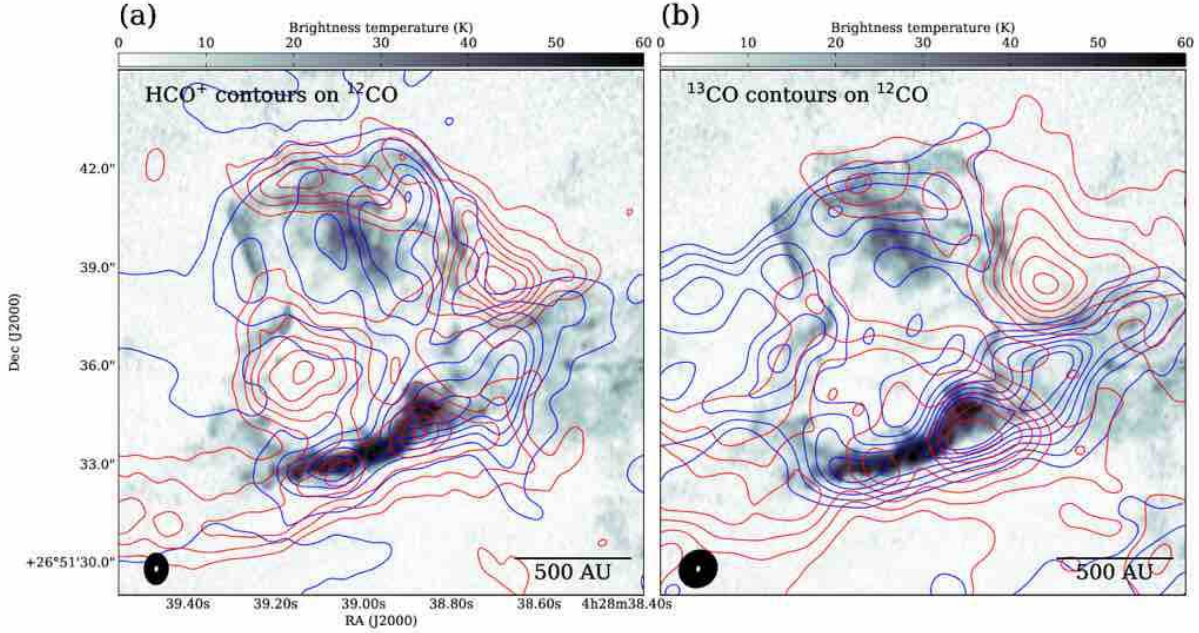}
\caption{Spatial distributions of HCO$^{+}$, $^{13}$CO, and $^{12}$CO in MC27/L1521F. (a), (b) Grayscale images show the peak brightness temperature maps of $^{12}$CO ($J$ = 3--2), same as Figure \ref{fig:bright}. The angular resolutions of the grayscale images and the contour images are given by white and black ellipses, respectively, in the lower left corners in each panel. (a) Blue and red contours show the image of the velocity-integrated intensity of HCO$^+$ ($J$= 3--2) obtained by the 12 m array alone with a range of 5.8--6.3 km\,s$^{-1}$ and 6.6--7.1 km\,s$^{-1}$. The lowest contour level is 0.24 K\,km\,s$^{-1}$. The subsequent contour levels of blue contours and red ones are 1.20  K\,km\,s$^{-1}$ and 0.40 K\,km\,s$^{-1}$, respectively. (b) Blue and red contours show the image of the velocity-integrated intensity of $^{13}$CO ($J$= 2--1) obtained by the 12 m array alone with the same velocity range used for (a). The lowest contour and subsequent contour step are 0.15 K\,km\,s$^{-1}$ and 0.20 K\,km\,s$^{-1}$, respectively. 
 \label{fig:Comp}}
\end{figure}
We argued that the turbulent field in this core is originated from the internal motions within the cloud core. 
However, we note that we cannot exclude external origins of the turbulent velocity field, such as gas feeding from the surrounding environment. \cite{Takakuwa11} found that the molecular abundance of carbon-chain molecules (e.g., C$_4$H, CH$_3$CCH) in this object is much higher than that of IRAM 04191+1522, a very low-luminosity protostar with a similar evolutionary status of MC27/L1521F \citep{Andre99,Dunham06}. They discussed that the discrepancies between the two objects are caused by the differences in their environments. MC27/L1521F is located in the central region of the Taurus molecular complex \cite[e.g.,][]{Mizuno95}, while IRAM 04191+1522 is isolated from that. In addition, the multiple velocity components are overlapped at the center of MC27/L1521F, as shown in Sec. \ref{subsec:1318}. A large amount of surrounding gas around MC27/L1521F can supply sources to enhance the abundance of the chemically young species \citep{Takakuwa11}. The high-density condensations are slightly redshifted from the systemic velocity of this core (Sec. \ref{subsec:1318}), indicating that the asymmetric high-density parts in the spatial/velocity space may be formed by perturbations from the redshifted 7.4 km\,s$^{-1}$ gas.  
Although distributions of CO and HCO$^+$ isotopes tend to avoid intensity peaks of dust emissions in cold and dense environments due to the molecular depletion \citep[e.g.,][]{Caselli99,Caselli02}, our observations do not show such evidence in the high-density condensations, MMS-2 and 3 (Figure \ref{fig:1318}, see also Papers I, and II).
It means that the condensations may be products from such external chemically young gas. 
\cite{Maureira17} also suggested that a complex velocity field in a very young protostellar core, Per-bolo 58, has two possible origins: infall motions from an inhomogeneous and flattened envelope (internal origin), and a consequence of accretion from large-scale filamentary structures (external origin). \\
\ The origin of warm gas in protostellar sources can be explained by shock heating by the protostellar outflow and/or jet if they are visible in CO or other molecular tracers. In the case of other Class 0 sources, high-resolution observations with ALMA revealed that warm gas ($>$50 K) components are seen with CO lines along the outflow directions and the cavity edges \citep{Lee14,Podio15,Yen17}. Although a bipolar nebula with the size scale of $>$1,000 au in MC27/L1521F is also seen with the {\it Spitzer} observations \citep{Bourke06,Terebey09}, the molecular distributions revealed in the CO and the HCO$^{+}$ failed to trace the bipolar nebula (Papers I, and II). 
We also discussed in Paper III the possibility that the bipolar nebula could be an evidence of the past outflow activity instead of the current one and suggested that the mass accretion onto the protostar/disk has halted quite recently.
We reported CO/HCO$^{+}$ high-velocity clouds as the current outflow with the size scale of a few hundred au (Papers I, and II) which corresponds to the compact structure seen at 1.9--2.75 km\,s$^{-1}$ in Figure \ref{fig:chmapCO}. There is a possibility that the compact structure is not outflowing gas because its shape is quite similar to the thin filaments (Sec. \ref{subsec:COfilament}) and connecting to the Bright filament rather than the protostar. 
Even if the component is a true outflow, the extremely tiny one is not enough to heat up the wide regions where we see the warm CO gas.
However, there is a possibility that the past outflow activities are related to the formation of the inhomogeneous warm gas. 
Thermal instability behind the shock layer \citep[e.g.,][]{Koyama00, Koyama02} can be a candidate mechanism to promote such kind of complex structure formation. According to calculations by \cite{Aota13}, colliding flows with the velocity of 10 km\,s$^{-1}$ can produce a thermally unstable layer in high-density (10$^4$ cm$^{-3}$) environments. The past outflow activities may be an origin of such high-velocity gas. 
The resultant unstable wavelength induced by the thermal instability ranges from $\sim$20 to 2,000 au depending on the preshock density and shock speed \citep{Aota13}. The timescale of complex structure formation by the thermal instability is given by the cooling time, which can be as short as $\sim$10$^3$ years inside a cloud after the shock compression \citep{Koyama00}. The temperature of the structures decreases down to the preshock temperature of $\sim$10 K after a few cooling time, indicating that warm gas structures can remain only for a short period ($\sim$10$^3$--10$^4$ years) after their creation. Thus, there is a possibility that some of the observed warm structures represent remnants of the past outflow activity that was halted very recently. 
\\

\subsection{Formation of protostar and dense core in a turbulent environment \label{subsec:formation}}
In this subsection, we discuss the formation of the protostar and the starless condensations in relation to the internal gas kinematics in MC27/L1521F. The gravitational force predicted from the mass of the protostar ($\sim$0.2 $M_{\odot}$, Paper III) and the highly concentrated column density distributions (Paper II) of the parental core is considered to be strong enough to accumulate the surrounding gas on to the protostar. However, the central velocities of the protostar (MMS-1) and the high-density condensations (MMS-2,3) are apparently shifted from that of the parental core with the relative velocity difference of $\sim$0.2 km\,s$^{-1}$ (Sec. \ref{subsec:1318}), which is corresponding to a sound speed at a temperature of 10 K. This fact indicates that the protostar and the condensations were formed under the effect of not only the self-gravity but the complex velocity field in the core. 
Numerical simulations done by \cite{Goodwin04} proposed that turbulence is a fundamental ingredient of core dynamics and leads to the production of a multiple star system. Our previous studies also proposed that the initial condition of low-mass (multiple) star formation is highly dynamical based on the complex gas distributions in MC27/L1521F (Papers I, and II). The presence of the warm gas implies that shock compressions in small scales created the seeds of protostars. The densities and the total masses of the starless condensations, MMS-2 and 3, are 10$^6$--10$^7$ cm$^{-3}$ and 10$^{-4}$--10$^{-3}$ $M_{\odot}$, respectively (Table \ref{tableParams}, see also Papers I, and II). The estimated masses are much smaller than the Jeans mass of the densities, indicating that the condensations are not formed by gravitational instability. Such tiny prestellar cores can be formed by turbulent shocks predicted by frameworks of formation of brown dwarfs \citep[e.g.,][]{Padoan04,Whitworth07}. This scenario can be a good candidate to give a convincing interpretation of the existence of the high-density condensations in MC27/L1521F. The condensations are mainly located on the western side of the Bright filament and Thin filament A, possible shocked layers, as shown in Figure \ref{fig:bright}. This fact implies that the high-density regions are created by the shock compressions. On the other hand, considering the fact that there is no significant mass accretion onto the protostar in spite of the existence of a large amount of surrounding gas (Paper III), the turbulent motions also might disturb the accretion activity. 
In this system, the gas sweeping by turbulent flow might play an important role in forming the compact and detached disk with the size scale of $\sim$10 au. The present observations may provide a scenario in which turbulence motions play important roles in constraining the stellar mass as well as the disk mass/size and in forming substellar mass condensations. \\
\ The remaining issues are when and how complex gas structures are formed due to turbulent motions. 
Internal turbulences of dense cores is mostly dissipated until the beginning of the gravitational contraction (c.f. $``$coherent core,'' \citealt{Goodman98}). 
 \cite{Balle17} reported that a large velocity dispersion can be created by a hierarchical collapse of an inhomogeneous 1,000 $M_{\odot}$ massive core.
However, simulations of low-mass ($\sim$7$-$40 $M_{\odot}$) protostellar collapse of initially inhomogeneous clouds with turbulence and magnetic fields \citep{Matsumoto11} provided a smooth isodensity surface in the high-density region created by the gravitational collapse. 
 Actually, according to the Larson'{}s law \citep{Larson81}, the velocity dispersion for the length scale studied in this paper ($<$10$^3$ au) corresponds to $<$ 0.1 km\,s$^{-1}$, which cannot result in a significant shock heating.
The complex internal substructures of dense cores at an early phase of gravitational contraction (i.e., before protostellar core formation) tend to be unexpected in theoretical aspects.  \cite{Matsumoto15a} also suggested that the internal turbulence is developed after the protostar formation due to the gravitational interactions between the protostar and the envelope gas. However, single-dish observations of prestellar cores \citep[e.g., ][]{Caselli02} detected some hints of complex gas motions. Recent ALMA observations have been revealing internal substructures of starless phases (\citealt{Ohashi18}; K. Tachihara et al. in prep.). As an alternative interpretation, we mentioned that the internal complex gas and the high-density condensations may have originated as a convergence of external gas (Sec. \ref{subsec:originshock}). Such effects are considered to happen disregarding the protostar formation incidentally. Although we cannot exactly conclude the origin of turbulent motion and protostar formation so far, future ALMA surveys with similar observational settings of this study toward prestellar cores and protostellar cores in various environments will provide us with crucial information to understand the initial condition of (multiple) star formation.

\section{Summary \label{sec:summary}}
\ We have presented ALMA Cycle 3 observations toward one of the densest cores MC27/L1521F, which contains a very low-luminosity protostar with the stellar mass of 0.2 $M_{\odot}$ and a few starless condensations. Our main results are summarized as follows.\\
1. High-angular resolution observations ($\sim$0\farcs2) in $^{12}$CO ($J$= 3--2) have revealed warm ($>$15--60 K) CO filamentary clouds and tiny CO clumps with the sizes of a few tens of au to $\sim$1,000 au. Although some hints of the warm gas were previously obtained by single-dish observations in high-$J$ CO lines \citep{Shinnaga09}, we identified small-scale warm structures in the cold protostellar core for the first time. \\
2. Detailed gas distributions with the density range of 10$^3$--10$^5$ cm$^{-3}$ around the protostar and the starless condensations are revealed in $^{13}$CO ($J$ = 2--1) and C$^{18}$O ($J$ = 2--1) by combining the Main array (12 m array) with the ACA (7 m+TP array) data. The interferometric observations revealed that high-density parts associated with the protostar and dense core formation with arc-like morphologies are slightly redshifted from the systemic velocity of the parental core. This fact indicated that the multiple density components are inhomogeneously distributed in both the velocity and space.
\\
3. We suggest that the complex warm CO gas can be formed by shocks induced by interactions among the different velocity/density gases that originated from the turbulent motions, and/or a consequence of the past outflow activities. The $^{13}$CO and HCO$^{+}$ envelope gases also indicate that different velocity components may interact with each other to form the shock-heated layers seen in the bright CO emission. 
Our present studies suggested that warm CO distributions can be a powerful indicator to characterize turbulent nature and structure formations, such as protostar/disk and dense condensations, in an early phase of star formation.\\
\ Such small-scale turbulent motions are not expected in a core following the Larson'{}s law. Thus how such a fast turbulent flow survives in this very dense medium is unknown and remains to be studied.
We also finally remark that shock heating directly represents dissipation of turbulent energies. Recent MHD simulations suggest that a large fraction of the turbulent energy dissipates in shock waves \citep{Lehmann16}. \cite{Tachihara00} predicted that high-$J$ transitions of CO lines can be detected at the shock-heated regions by collisions of turbulent eddies. Our ALMA observations of MC27/L1521F may be glimpsing a moment of the turbulence dissipation. If the dissipation process of the turbulence is indirectly seen as such high-temperature CO gas, similar structures can be also expected at non-starforming places dominated by turbulent motions.  K. Saigo et al. (in preparation) also found a high-temperature shell-like gas, which is possibly generated by turbulent flow, at a few thousand au away from a protostellar source by the $^{12}$CO ($J$ = 2--1) ALMA observations in B59 region. ALMA observations with a resolution of a few tens of au in CO lines can be future subjects to understand a comprehensive nature among turbulent dissipation, star formation, and warm CO gas.

\acknowledgments
This paper makes use of the following ALMA data: ADS/ JAO.ALMA\#2011.0.00611.S, 2012.1.00239.S, and 2015.1.00340.S. ALMA is a partnership of the ESO, NSF, NINS, NRC, NSC, and ASIAA. The Joint ALMA Observatory is operated by the ESO, AUI/NRAO, and NAOJ. This work was supported by NAOJ ALMA Scientific Research Grant Numbers 2016-03B and JSPS KAKENHI (Grant No. 22244014, 23403001, 26247026, and 18K13582).

\begin{figure}[htbp]
\plotone{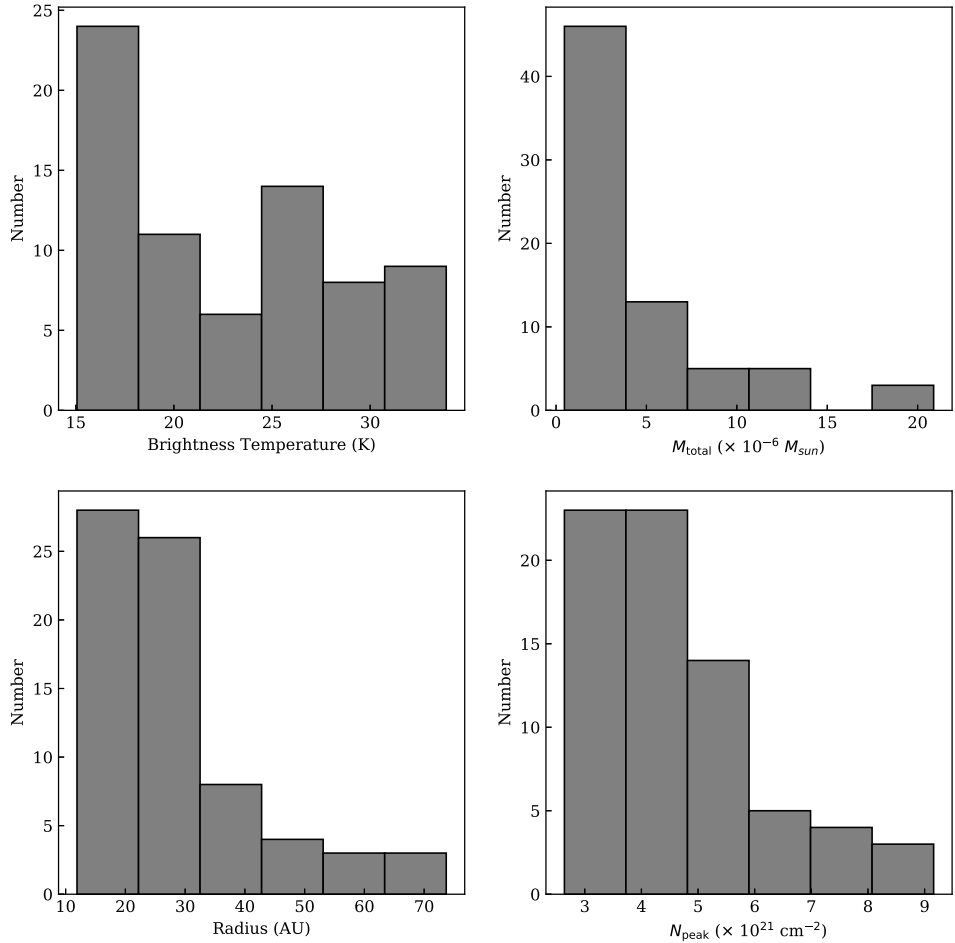}
\caption{Histograms of the physical properties of tiny CO clumps in MC27/L1521F. \label{fig:histogram}}
\end{figure}

\appendix
 The physical parameters of the tiny CO clumps identified by the dendrogram analysis (see Sec. \ref{subsec:tiny}) are shown in Table \ref{table:COclumps} and Figure \ref{fig:histogram}. We also estimated the column density and the total mass of the tiny clumps as reference values by using a standard Galactic CO conversion factor, 2.0 $\times$ 10$^{20}$ (K\,km\,s$^{-1}$)$^{-1}$\,cm$^{-2}$ \citep[e.g.,][]{Dame01} with an assumption of CO ($J$ = 3--2)/CO ($J$ = 1--0) intensity ratio of 1.  Although the application of the conversion factor derived by large-scale CO observations is not trivial, the estimated column densities with a spatial resolution of 1\arcsec are roughly corresponding to those from the $^{13}$CO observations (see Sec. \ref{subsec:COfilament}, and Sec. \ref{subsec:1318}) except for the Bright filament. The averaged radius, aspect ratio, peak column density ($N_{\rm peak}$), and total mass ($M_{\rm total}$) are 28 au, 2.0, 5 $\times$ 10$^{21}$ cm$^{-2}$, and 5 $\times$ 10$^{-6}$ $M_{\odot}$, respectively.

\begin{longtable}{lcccccccc}
\caption{Physical properties of tiny CO clumps \label{table:COclumps} }\\
\hline
Clump id &  $\alpha_{J2000.0}$ & $\delta_{J2000.0}$ & $T_{\rm peak}$ &  CO Luminosity                         &  Radius  &  Aspect ratio &  $N_{\rm peak}$                           &  $M_{\rm total}$  \\
{}             &  (degree)                  & (degree)                  &                    (K) & (K\,km\,s$^{-1}$\,arcsec$^{2}$) &       (AU) &  {}                 &  $\times$ 10$^{21}$ (cm$^{-2}$) &  $\times$ 10$^{-6}$ ($M_{\odot}$) \\
\hline
0  &  67.1615 &  26.85965 &        17.2 &                    0.52 &         16.6 &           1.2 &                    4.5 &                  1.0 \\
1  &  67.1616 &  26.85963 &        15.1 &                    0.47 &         15.3 &           1.5 &                    4.2 &                  0.9 \\
2  &  67.1618 &  26.85975 &        33.4 &                    0.60 &         12.1 &           1.9 &                    5.7 &                  1.2 \\
3  &  67.1603 &  26.85949 &        16.9 &                    1.19 &         25.0 &           3.0 &                    2.9 &                  2.4 \\
4  &  67.1635 &  26.85953 &        26.4 &                    1.94 &         25.6 &           1.5 &                    4.5 &                  3.9 \\
5  &  67.1600 &  26.85956 &        15.5 &                    0.23 &         11.9 &           1.7 &                    2.6 &                  0.5 \\
6  &  67.1601 &  26.85961 &        17.5 &                    1.49 &         27.4 &           1.2 &                    3.0 &                  3.0 \\
7  &  67.1614 &  26.85961 &        31.5 &                    1.48 &         20.1 &           1.3 &                    5.4 &                  2.9 \\
8  &  67.1598 &  26.85969 &        17.9 &                    1.26 &         28.5 &           1.8 &                    3.0 &                  2.5 \\
9  &  67.1601 &  26.85972 &        17.1 &                    0.69 &         18.6 &           1.4 &                    2.9 &                  1.4 \\
10 &  67.1600 &  26.85985 &        15.7 &                    2.57 &         37.3 &           1.9 &                    4.5 &                  5.1 \\
11 &  67.1598 &  26.85979 &        16.5 &                    0.31 &         14.0 &           1.7 &                    2.8 &                  0.6 \\
12 &  67.1617 &  26.86019 &        25.0 &                    9.75 &         54.4 &           2.3 &                    8.9 &                 19.4 \\
13 &  67.1633 &  26.86031 &        28.2 &                    3.71 &         30.6 &           2.6 &                    9.2 &                  7.4 \\
14 &  67.1632 &  26.86044 &        25.7 &                    1.08 &         17.6 &           2.0 &                    8.0 &                  2.1 \\
15 &  67.1616 &  26.86071 &        22.4 &                    1.21 &         21.0 &           2.8 &                    6.3 &                  2.4 \\
16 &  67.1616 &  26.86085 &        23.4 &                    0.82 &         17.3 &           2.8 &                    4.0 &                  1.6 \\
17 &  67.1618 &  26.86085 &        22.8 &                    1.28 &         19.5 &           1.8 &                    7.5 &                  2.5 \\
18 &  67.1620 &  26.86118 &        24.6 &                    4.33 &         43.5 &           2.6 &                    6.1 &                  8.6 \\
19 &  67.1636 &  26.85906 &        26.0 &                    3.61 &         30.3 &           1.4 &                    7.6 &                  7.2 \\
20 &  67.1633 &  26.85927 &        26.2 &                    1.64 &         22.8 &           2.6 &                    4.5 &                  3.3 \\
21 &  67.1637 &  26.85939 &        15.1 &                    0.44 &         15.6 &           2.2 &                    3.8 &                  0.9 \\
22 &  67.1615 &  26.85959 &        27.4 &                    1.02 &         17.3 &           1.3 &                    4.7 &                  2.0 \\
23 &  67.1613 &  26.86001 &        16.3 &                    2.76 &         38.2 &           1.3 &                    4.7 &                  5.5 \\
24 &  67.1603 &  26.86014 &        15.8 &                    4.77 &         55.2 &           2.1 &                    3.9 &                  9.5 \\
25 &  67.1600 &  26.86043 &        15.3 &                    6.99 &         71.3 &           1.6 &                    3.7 &                 13.9 \\
26 &  67.1624 &  26.86083 &        33.9 &                    2.39 &         23.5 &           1.7 &                    5.8 &                  4.8 \\
27 &  67.1616 &  26.86098 &        31.1 &                    1.81 &         21.7 &           1.8 &                    5.3 &                  3.6 \\
28 &  67.1631 &  26.86109 &        19.3 &                    3.31 &         38.2 &           2.1 &                    5.5 &                  6.6 \\
29 &  67.1616 &  26.86110 &        33.2 &                    2.33 &         24.4 &           2.4 &                    5.6 &                  4.6 \\
30 &  67.1624 &  26.86133 &        29.1 &                    1.00 &         16.7 &           2.0 &                    5.0 &                  2.0 \\
31 &  67.1626 &  26.86135 &        33.7 &                    2.17 &         23.8 &           3.5 &                    5.7 &                  4.3 \\
32 &  67.1632 &  26.86138 &        30.5 &                    1.18 &         18.0 &           1.7 &                    5.2 &                  2.4 \\
33 &  67.1631 &  26.86140 &        29.2 &                    2.14 &         24.4 &           1.7 &                    5.0 &                  4.3 \\
34 &  67.1634 &  26.85899 &        26.7 &                    1.16 &         19.4 &           1.6 &                    4.5 &                  2.3 \\
35 &  67.1635 &  26.86123 &        17.6 &                    1.55 &         28.6 &           1.5 &                    3.0 &                  3.1 \\
36 &  67.1635 &  26.86138 &        19.1 &                    1.04 &         22.3 &           1.5 &                    3.3 &                  2.1 \\
37 &  67.1630 &  26.86152 &        30.3 &                    2.20 &         24.8 &           2.5 &                    5.2 &                  4.4 \\
38 &  67.1629 &  26.86154 &        30.2 &                    1.24 &         18.9 &           2.2 &                    5.1 &                  2.5 \\
39 &  67.1624 &  26.86162 &        20.4 &                    3.88 &         42.1 &           2.4 &                    3.5 &                  7.7 \\
40 &  67.1625 &  26.86169 &        19.7 &                    1.71 &         28.3 &           1.2 &                    3.3 &                  3.4 \\
41 &  67.1631 &  26.86169 &        21.0 &                    2.09 &         30.9 &           2.5 &                    3.6 &                  4.1 \\
42 &  67.1619 &  26.86169 &        24.8 &                    1.52 &         24.4 &           1.1 &                    4.2 &                  3.0 \\
43 &  67.1622 &  26.86176 &        19.4 &                    1.16 &         23.0 &           1.7 &                    3.3 &                  2.3 \\
44 &  67.1620 &  26.86176 &        19.6 &                    1.25 &         22.9 &           1.7 &                    3.3 &                  2.5 \\
45 &  67.1633 &  26.86178 &        15.5 &                    0.50 &         17.3 &           2.1 &                    2.6 &                  1.0 \\
46 &  67.1633 &  26.86004 &        17.1 &                    1.30 &         34.1 &           1.8 &                    3.9 &                  2.6 \\
47 &  67.1634 &  26.86142 &        15.9 &                    1.01 &         26.0 &           2.0 &                    4.0 &                  2.0 \\
48 &  67.1639 &  26.85950 &        20.1 &                    6.90 &         73.7 &           1.5 &                    4.6 &                 13.7 \\
49 &  67.1598 &  26.85975 &        23.7 &                    1.36 &         22.2 &           1.1 &                    4.0 &                  2.7 \\
50 &  67.1600 &  26.85983 &        27.4 &                    5.01 &         43.1 &           1.9 &                    4.7 &                 10.0 \\
51 &  67.1615 &  26.85989 &        27.3 &                   10.27 &         65.5 &           2.3 &                    7.0 &                 20.4 \\
52 &  67.1611 &  26.85974 &        15.6 &                    5.67 &         50.2 &           2.5 &                    4.7 &                 11.3 \\
53 &  67.1597 &  26.85994 &        29.9 &                    3.00 &         29.8 &           1.9 &                    8.9 &                  6.0 \\
54 &  67.1602 &  26.85997 &        19.3 &                    1.57 &         27.9 &           1.7 &                    3.3 &                  3.1 \\
55 &  67.1600 &  26.86000 &        19.8 &                    1.27 &         23.3 &           3.1 &                    3.4 &                  2.5 \\
56 &  67.1606 &  26.86000 &        31.6 &                    1.91 &         23.9 &           1.2 &                    5.4 &                  3.8 \\
57 &  67.1599 &  26.85999 &        26.0 &                    1.04 &         18.4 &           1.2 &                    4.4 &                  2.1 \\
58 &  67.1605 &  26.85998 &        23.2 &                    0.47 &         12.7 &           1.9 &                    3.9 &                  0.9 \\
59 &  67.1604 &  26.86018 &        23.9 &                    1.85 &         29.0 &           2.0 &                    6.1 &                  3.7 \\
60 &  67.1607 &  26.86022 &        17.9 &                    2.44 &         35.0 &           1.9 &                    3.0 &                  4.9 \\
61 &  67.1598 &  26.86029 &        17.9 &                    1.32 &         27.6 &           2.3 &                    3.0 &                  2.6 \\
62 &  67.1604 &  26.86035 &        27.3 &                    5.78 &         49.8 &           2.3 &                    6.4 &                 11.5 \\
63 &  67.1608 &  26.86039 &        31.8 &                   10.50 &         60.0 &           2.0 &                    7.0 &                 20.9 \\
64 &  67.1611 &  26.86035 &        17.9 &                    0.72 &         18.8 &           1.7 &                    3.0 &                  1.4 \\
65 &  67.1605 &  26.86059 &        19.7 &                    3.10 &         38.9 &           1.9 &                    3.4 &                  6.2 \\
66 &  67.1636 &  26.86096 &        32.4 &                    5.79 &         40.1 &           5.4 &                    5.5 &                 11.5 \\
67 &  67.1637 &  26.86121 &        27.1 &                    1.39 &         20.4 &           3.3 &                    4.6 &                  2.8 \\
68 &  67.1636 &  26.86120 &        27.7 &                    1.38 &         20.4 &           2.8 &                    4.7 &                  2.8 \\
69 &  67.1635 &  26.86127 &        16.6 &                    0.47 &         15.9 &           2.7 &                    2.8 &                  0.9 \\
70 &  67.1629 &  26.86155 &        17.6 &                    0.47 &         14.8 &           1.3 &                    3.0 &                  0.9 \\
71 &  67.1629 &  26.86160 &        17.7 &                    0.38 &         13.5 &           1.2 &                    3.0 &                  0.7 \\
\hline
\end{longtable}

\end{document}